\documentclass[10pt,onecolumn,a4paper]{IEEEtran}
\usepackage{amsmath}
\usepackage{graphicx}
\renewcommand{\baselinestretch}{1.3}

\newcommand{\figuramedia}[3]
{
\begin{figure}
  \centering
 \includegraphics[width=12cm]{#1}
  \caption{#2}\label{#3}
\end{figure}
}

\newcommand{\figura}[3]
{
\begin{figure}
  \centering
 \includegraphics[width=12cm]{#1}
  \caption{#2}\label{#3}
\end{figure}
}
\title{Saturation Throughput Analysis of IEEE 802.11 in Presence of Non Ideal Transmission Channel and Capture Effects}
\author{\authorblockN{F. Daneshgaran, M. Laddomada, F. Mesiti, M.
Mondin, and M. Zanolo \thanks{This work has been partially
supported by Euroconcepts, S.r.l. (http://www.euroconcepts.it) and
ISMB through the European IST project, OBAN
(http://oban.prz.tu-berlin.de/index.html), IST 6FP Contract No.
001889.}
\thanks{\textbf{To appear on IEEE Transactions on Communications, 2008}}
\thanks{F. Daneshgaran is with ECE Dept., CSU,
Los Angeles, USA.}
\thanks{M. Laddomada (\textrm{laddomada@polito.it}), F. Mesiti, M. Mondin and M. Zanolo
are with DELEN, Politecnico di Torino, Italy.}}}
\begin{document}
\maketitle

\begin{abstract}
In this paper, we provide a saturation throughput analysis of the
IEEE 802.11 protocol at the data link layer by including the
impact of both transmission channel and capture effects in
Rayleigh fading environment. Impacts of both non-ideal channel and
capture effects, specially in an environment of high interference,
become important in terms of the actual observed throughput. As
far as the 4-way handshaking mechanism is concerned, we extend the
multi-dimensional Markovian state transition model characterizing
the behavior at the MAC layer by including transmission states
that account for packet transmission failures due to errors caused
by propagation through the channel. This way, any channel model
characterizing the physical transmission medium can be
accommodated, including AWGN and fading channels. We also extend
the Markov model in order to consider the behavior of the
contention window when employing the basic 2-way handshaking
mechanism.

Under the usual assumptions regarding the traffic generated per
node and independence of packet collisions, we solve for the
stationary probabilities of the Markov chain and develop
expressions for the saturation throughput as a function of
the number of terminals, packet sizes, raw channel error rates,
capture probability, and other key system parameters. The theoretical
derivations are then compared to simulation
results confirming the effectiveness of the proposed models.
\end{abstract}
\clearpage
\section{Introduction}
Wireless Local Area Networks (WLANs) combine data connectivity
with user mobility. The Institute of Electrical and Electronic
Engineering (IEEE) 802.11 series of standards define both an
infrastructure mode, with at least one central Access Point (AP)
connected to a wired network, and an ad-hoc or peer-to-peer mode,
in which a set of wireless stations communicate directly with
another one without needing a central access point or wired
network connection.

WLANs have experienced an exponential growth in the recent past
[1-20]. The fundamental impetus has been to replicate the enormous
success of wired LANs with obvious advantages a wireless paradigm
would bring. The first series of standards by IEEE was ratified in
1997 but had relatively low data rates (1 or 2 Mbps). Two high
rate versions, namely, the IEEE802.11a and IEEE802.11b were later
ratified in 1999 and they have found widespread acceptance and
use. The enormous success of these standards has prompted multiple
working groups to extend some aspects of the basic protocol.

While in the ensuing presentation we shall focus on the basic
IEEE802.11b protocol, the analysis is easily extensible to other
versions of the basic protocol employing the same Medium Access
Control (MAC) mechanisms.

The MAC is central to embedding Quality of Service (QoS) features
and has two sub-layers \cite{standard_DCF_MAC}. The Distributed
Coordination Function (DCF) and Point Coordination Function (PCF).
PCF is generally a complex access method that can be implemented
in an infrastructure network. DCF is similar to Carrier Sense
Multiple Access with Collision Avoidance (CSMA/CA) method and is
further explained in the following.

With this background, let us provide a quick survey of the recent
literature related to the problem addressed in this paper. This
survey is by no means exhaustive and is meant to simply provide a
sampling of the literature in this fertile area.

The most relevant work to what is presented here is
\cite{Bianchi}, where the author
provided an analysis of the saturation throughput
of the basic 802.11 protocol assuming a two dimensional Markov
model at the MAC layer, making several fundamental assumptions: 1)
the mobile stations always have something to transmit (i.e., the
saturation condition), 2) there are no hidden terminals and there
is no capture effect (i.e., a terminal which perceives a higher
signal-to-noise ratio (SNR) relative to other terminals captures
the channel \cite{capture,zorzi_rao,Spasenovski} and limits access
to other terminals), 3) at each transmission attempt and
regardless of the number of retransmissions suffered, each packet
collides with constant and independent probability, and 4) the
transmission channel is ideal and the only cause of packet errors
are collisions. Clearly, the second and fourth assumption is not
valid in any real setting, specially when there is mobility and
when the transmission channel suffers from fading effects.

In \cite{Randhawa}, the authors extend the work of Bianchi to
multiple queues with different contention characteristics in the
802.11e variant of the standard with provisions of QoS. In
\cite{kong}, the authors present an analytical model, in which
most new features of the Enhanced Distributed Channel Access
(EDCA) in 802.11e such as virtual collision, different arbitration
interframe space (AIFS), and different contention windows are taken
into account. Based on the model, the throughput performance of
differentiated service traffic is analyzed and a recursive method
capable of calculating the mean access delay is presented. Both
articles referenced assume the transmission channel to be ideal.

In \cite{Qiao}, the authors look at the impact of channel induced
errors and the received SNR on the achievable throughput in a
system with rate adaptation whereby the transmission rate of the
terminal is adapted based on either direct or indirect
measurements of the link quality. In \cite{Chatzimisios}, the
authors deal with the extension of Bianchi's Markov model in order
to account for channel errors.

In this article we extend the previous work on the subject as
exemplified by the articles referenced above by looking at all the
three issues outlined before together. Our assumptions are
essentially similar to those of Bianchi \cite{Bianchi} with the
exception that we do assume the presence of both channel induced
errors and capture effects due to the transmission over Rayleigh
fading channels. Simulation results confirm the effectiveness of
the proposed models. As a reference standard, we use network
parameters belonging to the IEEE802.11b protocol, even though the
proposed mathematical models hold for any flavor of the IEEE802.11
family.

The rest of the article is organized as follows. Section~II
reviews the functionalities of the contention window procedure at
MAC layer. Section~III extends the Markov model initially proposed
by Bianchi, presenting modifications that account for transmission
errors and capture effects over Rayleigh fading channels employing
the 4-way handshaking technique. This section also provides an
expression for the saturation throughput of the link. In
section~IV we briefly extend the analysis of the two dimensional
Markov chain for the basic 2-way handshaking mechanism. In
section~V we present simulation results where typical MAC layer
parameters for IEEE802.11b are used to obtain throughput values as
a function of various system level parameters, capture
probability, and SNR. Section~VI is devoted to conclusions.
\section{Development of the Markov Model}
In this section, we present the rationales at the basis of the
proposed bi-dimensional Markov model useful for evaluating the
throughput of the DCF under the assumptions of finite number of
operating terminals, channel error conditions, capture effects in
Rayleigh fading transmission scenario, and packet transmission
based on the four-way handshaking access mechanism. For
conciseness, we will limit our presentation to the ideas needed
for developing the proposed model.
We invite the interested readers to refer to
\cite{standard_DCF_MAC,Bianchi} for many details on the operating
functionalities of the DCF.
\subsection{Markovian Model Characterizing the MAC Layer: Perfect Transmission Channel}
In \cite{Bianchi}, a discrete-time bi-dimensional Markov process
$(s(t),b(t))$ is presented for the computation of the throughput
of a WLAN using the IEEE 802.11 DCF under ideal channel
conditions.

It is based on two key assertions:
\begin{enumerate}
\item the probability $\tau$ that a station will attempt
transmission in a generic time slot is constant across all time
slots;
\item the probability $P_{col}$ that any transmission experiences
a collision is constant and independent of the number of
collisions already suffered.
\end{enumerate}
%
%
The random process $b(t)$ is used to represent the backoff
counter of each station. Backoff counter is decremented at the
start of every idle backoff slot and when it reaches zero, the
station transmits and a new value for $b(t)$ is set. The value of
$b(t)$ after each transmission depends on the size of the
contention window from which it is drawn.
%
A second process $s(t)$ is defined, representing the size of the
contention window $(W_i=2^i W,~i=s(t))$ from which $b(t)$ is
drawn.

We recall that a backoff time counter is initialized depending on the
number of failed transmissions for the transmitted packet. It is
chosen from the range $[0,W_i-1]$ following a uniform Probability
Mass Function (PMF), where $W_i$ is the contention window size at the
backoff stage $i$. At the first transmission attempt (i.e., for
$i=0$), the contention window size is set equal to a minimum value
$W_0=W$, and the process $s(t)$ takes on the value $s(t)=i=0$.

The backoff stage $i$ is incremented in unitary steps after each
unsuccessful transmission up to the maximum value $m$, while the
contention window is doubled at each stage up to the maximum value
$CW_{max}=2^m W$.

The backoff time counter is decremented as long as the channel is
sensed idle and stopped when a transmission is detected. The
station transmits when the backoff time counter reaches zero.
%
%
%
\section{Markovian Model Characterizing the MAC Layer under Real
Transmission Channel and Capture Effects Using the 4-Way
Handshaking Mechanism}
The main aim of this section is to propose an effective
modification of the bi-dimensional Markov process proposed in
\cite{Bianchi} in order to account for channel error conditions
and capture effects over Rayleigh fading channel under the
hypothesis of employing the four-way handshaking access mechanism.
It is useful to briefly recall the four-way handshaking mechanism
for highlighting the hypotheses at the basis of the proposed
Markov model.

A station that wants to transmit a packet, waits until the channel
is sensed idle for a Distributed InterFrame Space (DIFS), follows
the backoff rules and then preliminarily transmits a short Request
To Send (RTS) frame. The receiving station, after having received
an RTS frame, responds with a Clear To Send (CTS) frame, after a
Short InterFrame Space (SIFS). The transmitting station is allowed
to transmit its packet only after having correctly received the
CTS frame. Both RTS and CTS frames carry the information about the
length of the packet to be transmitted. This information is
available to any listening station, so that they can update their
Network Allocation Vector (NAV) with the information about the
period of time in which the channel will be busy. Therefore, every
station, hidden from either the transmitting or the receiving
station, by detecting just one frame among the RTS and CTS frames,
can delay transmissions by an appropriate amount, thus avoiding
collisions.

On the basis of this assumption, collisions can only occur with
probability $P_{col}$ on RTS packets, while transmission errors,
due to the channel, can occur with probability $P_e$ only on the
data frames. Therefore, if a station is free to send a data frame
(that means that the transmitting station received the CTS frame),
it can start the data frame transmission. This frame is protected
from collision, but not from transmission errors. In this work, we
assume that collision and transmission error events are
statistically independents.

Furthermore, in mobile radio environment, it may happen that the
channel is captured by a station whose power level is stronger
than the other stations transmitting at the same time. This may be
due to relative distances and/or channel conditions for each user
and may happen whether or not the terminals exercise power
control. As a matter of fact, capture effect reduces the collision
probability on the channel, since the stations whose power level
at the receiver is very low due to path attenuation, shadowing and
fading, are considered as interferers from the AP, thus raising
the noise floor.

To account for these phenomena, our Markov chain model depicted in
Fig.~\ref{fig.chain} includes a Transmit (TX) state, identified by
$(i,-1),~\forall i=0,\ldots,m$, which is not considered in the
original work of Bianchi in \cite{Bianchi}. To reach the transmit state,
there should be no collision or a collision in which the desired station captures
the channel (hence, in effect no collision is detected even though there is one).

To simplify the analysis, we make the assumption that the impact
of the channel induced errors on the RTS, CTS and Acknowledgment
(ACK) packets are negligible because of their short length. This
is justified on the basis of the assumption that the bit errors
inflicting the transmitted data are independent of each other.
Hence, the packet or frame error rate, identified respectively
with the acronyms PER or FER, is a function of the packet length,
with shorter packets having exponentially smaller probability of
error compared to longer packets.

Notice that FER in general also depends on the forward error
correcting code employed at the physical layer
\cite{laddomada_3,laddomada_4}. However, for convolutional coding,
as optionally foreseen in the IEEE 802.11b standard
\cite{standard_DCF_MAC}, bit error rate (BER) is independent from
the code block size, and, as a consequence, FER assumes increasing
values for larger and larger data packet sizes. It is known
that the BER for convolutional codes depends on the distance
spectrum of the code \cite{proakis}-\cite{laddomada_2} which is
independent of the block size.

We note that with sufficient interleaving we can always ensure
that the errors inflicting individual bits in a data packet are
independent of each other. Since the length of the ACK, CTS and
RTS frames are typically less than 20 bytes, while the information
baring packets are typically several hundreds of bytes long, the
assumption is clearly justified. Including the possibility that
the ACK, RTS and CTS frames can themselves be erroneous unduly
complicates an already complex Markov chain with no clear
benefits.
%
%

Let us discuss the Markov model shown in Fig.~\ref{fig.chain}.
Similar to the model in \cite{Bianchi}, $(m+1)$ different backoff
stages are considered (this includes the zero-th stage). The
maximum contention windows (CW) sizes is then $CWmax = 2^mW$,
and the relation $W_i = 2^iW$ is used to define the $i^{th}$
contention window size. An RTS can be transmitted only in the
$(i,0)$ states, $\forall i=0,\ldots,m$, while data can only be
transmitted in the $(i,-1)$ states, $\forall i=0,\ldots,m$ upon a
successful reception of the CTS. If collision occurs, the backoff
stage is incremented, so that the new state can be $(i+1,k)$ with
probability $P_{col}/W_{i+1}$, since a uniform distribution
between the states in the same backoff stage is considered. We
consider capture as a subset of the event of a collision. In other
words, the capture event can happen in the presence of collision,
by allowing the transmitting station with the higher received
power level at the access point to capture the channel. In this
case, there is no collision and the Markov model transits into one
of the transmitting states $(i,-1)$ depending on the current
contention stage. If no collision occurs, a data frame can be
transmitted, and the transmitting station enters state $(i,-1)$,
based on the backoff stage it belongs to. From state $(i,-1)$, if
the transmission is successful, the transmitting station re-enters
the initial backoff stage, identified by $i=0$. Otherwise, if
errors occur during transmission, the ACK packet is not sent, an
ACK-timeout occurs, and the backoff stage is changed to $(i+1,k)$
with probability $P_e/W_{i+1}$, where $P_e$ is the packet error
probability for the employed physical layer.

The Markov Process of Fig. \ref{fig.chain} is governed by the
following transition probabilities\footnote{$P_{i,k|j,n}$ is short
for $P\{s(t+1)=i,b(t+1)=k|s(t)=j,b(t)=n\}$.}:
\begin{equation}\label{eq.process}\small
\begin{array}{lll}
P_{i,k|i,k+1} &= 1,                     &~ k \in [0,W_i-2], ~ i \in [0,m] \\
P_{i,-1|i,0}  &= 1-P_{col},   &~ k=0,~ i \in [0,m] \\
P_{0,k|i,-1}  &= (1-P_{e})/W_0,         &~ k \in [0,W_0-1], ~ i \in [0,m] \\
P_{i,k|i-1,0}   &= P_{col}/W_i,&~ k \in [0,W_i-1], ~ i \in [1,m] \\
P_{i,k|i-1,-1}&= P_{e}/W_i,&~ k \in [0,W_i-1], ~ i \in [1,m] \\
P_{m,k|m,0}   &= P_{col}/W_m,&~ k \in [0,W_m-1]  \\
P_{m,k|m,-1}   &= P_{e}/W_m,&~ k \in [0,W_m-1]
\end{array}
\end{equation}
The first equation in (\ref{eq.process}) states that, at the
beginning of each slot time, the backoff time is decremented. The
second equation accounts for the fact that after a successful RTS
transmission and CTS reception, the data transmission state is
$(s(t)=i,b(t)=-1)$ with probability $1-P_{col}$. The third
equation accounts for the fact that after a successful
transmission, a new packet transmission starts with backoff stage
0. The other equations deal with unsuccessful transmissions.
Capture effects are accounted for in the definition of the
collision probability. The fourth and sixth equations deal with
RTS frame collisions. In this case, a new backoff stage is
scheduled. Finally, the fifth and seventh equations model failed
transmissions due to channel induced errors. In this situation, a
new backoff stage is scheduled as well.
\subsection{Markovian Process Analysis and Throughput Computation}
Next line of pursuit consists in finding a solution of
the stationary distribution:
\[
b_{i,k}=\lim_{t\rightarrow \infty}P[s(t)=i,b(t)=k],~\forall
k\in[-1,W_i-1],~\forall i\in[0,m]
\]
that is the probability of a station occupying a given state at
any discrete time.

For the sake of simplifying the evaluation of the normalization
condition of the bidimensional Markov chain, let us express all
the probabilities $b_{i,k}$ as a function of $b_{0,0}$. To this
end, note the following relations:
\begin{equation}\label{trans_states_probabilities}
\begin{array}{ll}
b_{i,-1}=b_{i,0}\left(1-P_{col}\right),~\forall i\in
[0,m]&\\
b_{i,0}=b_{i-1,-1}P_e+b_{i-1,0}P_{col},~\forall i\in
[1,m)&\\
\end{array}
\end{equation}
while for $i=m$ we have:
\[
\begin{array}{ll}
b_{m,0}= b_{m-1,-1}P_e
+b_{m-1,0}P_{col}+&\\
+b_{m,-1}P_e+b_{m,0}P_{col}&
\end{array}
\]
from which it is possible to obtain:
\[
\begin{array}{ll}
b_{m,0}\left(1-P_{col}\right)= b_{m-1,-1}P_e
+b_{m,-1}P_e+&\\
+b_{m-1,0}P_{col}&
\end{array}
\]
These relations will be used later for solving the normalization
condition applied to the bi-dimensional Markov model shown in Fig.
\ref{fig.chain}.

Upon defining an equivalent probability of failed transmission,
$P_{eq}$\footnote{For simplicity, we assume that at each
transmission attempt any station will encounter a constant and
independent probability of failed transmission, $P_{eq}$,
independently from the number of retransmissions already suffered
from each station.}, that takes into account the need for a new
contention due to either RTS collision ($P_{col}$) or channel
errors affecting the data frame ($P_e$), i.e.,
\begin{equation}\label{eq.equ}
P_{eq}=P_{col}+P_e-P_e\cdot  P_{col}
\end{equation}
after some simple algebra involving the previous relations, it is
possible to obtain the following relations:
\begin{equation}\label{collision_probabilities}
\begin{array}{ll}
 b_{i,0}=P_{eq} b_{i-1,0},& ~\forall i\in [1,m)\\
 b_{m,0}=P_{eq} b_{m-1,0}+P_{eq} b_{m,0},&~i=m
\end{array}
\end{equation}
Proceeding backward, the following results follow by inspection:
\begin{equation}\label{collision_probabilities_2}
\begin{array}{ll}
b_{i,0}=P_{eq}^ib_{0,0},&~\forall i\in [1,m)\\
b_{m,0}=\frac{P_{eq}^m}{1-P_{eq}}b_{0,0},&~i=m
\end{array}
\end{equation}
The other stationary probabilities for any $k\in[1,W_i-1]$ follow
by resorting to the state transition diagram:
\begin{equation}\label{eq.bik}
b_{i,k} = \frac{W_i-k}{W_i}\left\{
\begin{array}{ll}
(1-P_{e})\sum_{j=0}^m b_{j,-1}, &~i=0\\
P_{col}b_{i-1,0}+P_{e}b_{i-1,-1}, &~i\in [1,m)\\
P_{col}(b_{m-1,0}+b_{m,0})+\\
+P_{e}(b_{m-1,-1}+b_{m,-1}), &~ i=m
\end{array} \right.
\end{equation}
Employing the normalization condition, after some mathematical
manipulations, it is possible to obtain:
%
%
\begin{equation}\small
\begin{array}{l}
1=\sum_{i=0}^m \sum_{k=-1}^{w_i-1} b_{i,k}= b_{0,0}
\frac{2(1-P_{eq})W(1-(2P_{eq})^{m-1})P_{eq}+
P_x(1-P_{col})\left((1-P_{e})(W+1)+2\right)+
P_x(1-P_{eq}^{m-1})P_{eq}+(2^mW+1)P_xP_{eq}^{m}}{2P_x(1-P_{eq})}
\end{array}
\end{equation}
whereby, $P_x=1-2P_{eq}$.
Normalization condition yields the following equation for
computation of $b_{0,0}$:
%
%
\begin{equation}\small
\begin{array}{ll}
b_{0,0}=\frac{2P_x(1-P_{eq})}{2(1-P_{eq})W(1-(2P_{eq})^{m-1})P_{eq}+
P_x(1-P_{col})\left((1-P_{e})(W+1)+2\right)+
P_x(1-P_{eq}^{m-1})P_{eq}+(2^mW+1)P_xP_{eq}^{m}}
\end{array}
\end{equation}
This result is then used to compute $\tau$, the probability that a
station starts a transmission in a randomly chosen slot time. In
fact, taking into account that an RTS transmission occurs when the
backoff counter reaches zero, we have:
%
%
\begin{equation}\label{eq.tau}
\begin{array}{ll}
\tau=\sum_{i=0}^m
b_{i,0}=\frac{b_{0,0}}{1-P_{eq}}=\frac{2P_x}{2(1-P_{eq})W(1-(2P_{eq})^{m-1})P_{eq}+
P_x(1-P_{col})\left((1-P_{e})(W+1)+2\right)+
P_x(1-P_{eq}^{m-1})P_{eq}+(2^mW+1)P_xP_{eq}^{m}}
\end{array}
\end{equation}
The collision probability, needed to compute $\tau$, can be found
considering that using a 4-way hand-shaking an RTS frame from a
transmitting station encounters a collision if in a time slot, at
least one of the remaining $(N-1)$ stations transmit
simultaneously an RTS frame, and there is no capture. In our
model, we assume that capture is a subset of the collision events.
This is indeed justified by the fact that there is no capture
without collision, and that capture occurs only because of
collisions between a certain number of transmitting stations
attempting to transmit simultaneously on the channel.
\begin{equation}\label{eq.col}
P_{col} = 1-(1-\tau)^{N-1}-P_{cap}
\end{equation}
As far as capture effects are concerned, we resort to the
mathematical formulation proposed in \cite{zorzi_rao,Spasenovski}.
In particular, under the hypothesis of power-controlled stations
in infrastructure mode, the capture probability conditioned on $i$
interfering frames can be defined as follows:
\begin{equation}\label{eq.capture_conditional}
P_{cp}\left(\gamma>z_o
g(S_f)|i\right)=\frac{1}{{\left[1+z_{0}g(S_{f})\right]}^{i}}
\end{equation}
whereby $\gamma$, defined as $P_u/\sum_{k=1}^{i}P_k$, is the ratio
of the power $P_u$ of the useful signal and the sum of the
powers of the $i$ interfering channel contenders transmitting
simultaneously $i$ frames, $g(S_f)$ is the inverse of the processing gain of the
correlation receiver, and $z_0$ is the capture ratio, i.e., the
value of the signal-to-interference power ratio identifying the
capture threshold at the receiver. Notice that
(\ref{eq.capture_conditional}) signifies the fact that capture
probability corresponds to the probability that the power ratio
$\gamma$ is above the capture threshold $z_{0}g(S_{f})$ which
considers the inverse of the processing gain $g(S_{f})$. For Direct Sequence
Spread Spectrum (DSSS) using a 11-chip spreading factor
($S_f=11$), we have $g(S_{f})=\frac{2}{3S_f}$.

Upon defining the probability of generating exactly $i+1$
interfering frames over $N$ contending stations in a generic slot
time:
\[
{N \choose i+1}\tau^{i+1}(1-\tau)^{N-i-1}
\]
the frame capture probability $P_{cap}$ can be obtained as follows:
\begin{equation}\label{capture_probability}
P_{cap}=\sum_{i=1}^{N-1}{N \choose
i+1}\tau^{i+1}(1-\tau)^{N-i-1}P_{cp}\left(\gamma>z_o
g(S_f)|i\right)
\end{equation}
Putting together equations (\ref{eq.equ}), (\ref{eq.tau}),
(\ref{eq.col}) and (\ref{capture_probability}), the following
nonlinear system can be defined and solved numerically, obtaining
the values of $\tau$ (defined in~(\ref{eq.tau})), $P_{col}$,
$P_{eq}$, and $P_{cap}$:
\begin{equation}\label{eq.system}
\left\{ \begin{array}{ll} \tau & \\
P_{col} &= 1-(1-\tau)^{N-1} -P_{cap}\\
P_{eq}   &=
P_{col}+P_e-P_e\cdot  P_{col}\\
P_{cap}&=\sum_{i=1}^{N-1}{N \choose
i+1}\tau^{i+1}(1-\tau)^{N-i-1}\frac{1}{{(1+z_{0}g(S_{f}))}^{i}}
\end{array} \right.
\end{equation}
The final step in the analysis is the computation of the
normalized system throughput, defined as the fraction of time the
channel is used to successfully transmit payload bits:
\begin{equation}\small
\label{eq.system2} S = \frac{P_t \cdot P_s\cdot
(1-P_e)E[PL]}{(1-P_t
)\sigma+P_t(1-P_s)T_c+P_tP_s(1-P_e)T_s+P_tP_sP_eT_e}
\end{equation}
where
\begin{itemize}
\item $P_t$ is the probability that there is at least one
transmission in the considered slot time, with $N$ stations
contending for the channel, each transmitting with probability
$\tau$:
\begin{equation}
P_t=1-(1-\tau)^N
\end{equation}

\item $P_s$ is the conditional probability that an RTS
transmission occurring on the channel is successful. This event
corresponds to the case in which exactly one station transmits in
a given slot time, or two or more stations transmit simultaneously
and capture by the desired station occurs. These conditions yields the following
probability:
\begin{equation}
P_s=\frac{N\tau(1-\tau)^{N-1}+P_{cap}}{P_t}
\end{equation}

\item $T_c$, $T_e$ and $T_s$ are the average time a  channel is
sensed busy due to an RTS collision, error affected data frame
transmission time and successful data frame transmission times,
respectively. Knowing the time durations for RTS, CTS, ACK frames,
CTS and ACK timeout, DIFS, SIFS, $\sigma$, data packet length ($PL$) and
PHY and MAC headers duration ($H$), and propagation delay $\tau_p$,
$T_c$, $T_s$, and $T_e$ can be computed as follows \cite{kong}:
\begin{equation}\label{tempi_coll_error}
 \begin{array}{ll}
T_c       &= RTS+CTS_{timeout}            \\
T_e       &= RTS+SIFS+\tau_p+CTS+SIFS+\tau_p+\\
&+H+PL+ACK_{timeout}  \\
T_s       &= RTS+SIFS+ \tau_p+CTS+SIFS+\tau_p+\\
&+H+PL+SIFS+\tau_p+ACK+DIFS+\tau_p  \\
\end{array}
\end{equation}
\item $E[PL]$ is the average packet payload length.

\item $\sigma$ is the duration of an empty slot time.
\end{itemize}
The setup described above is used in Section
\ref{Simulation_results_section} for DCF simulation at the MAC
layer.
\section{Analysis of the Markov chain for the basic 2-way handshaking mechanism in
presence of channel errors and capture effects}
This section briefly provides the extension of the Markov model
for the contention model to the case of using the basic two-way
handshaking mechanism.

Let us review the basic functionalities of the two-way handshaking
technique. No RTS/CTS exchange of packets is used.
After the successful reception of a data frame, the
receiver sends an ACK frame to the transmitter. Only upon a
correct ACK frame reception, the transmitter assumes successful
delivery of the corresponding data frame. If an ACK frame is
received in error or if no ACK frame is received, due possibly to
an erroneous reception of the preceding data frame, the
transmitter will contend again for the medium.

Based on these assumptions, using the two-way handshaking
mechanism there is no evident distinction between the transmission
of information and protocol data. In this respect, the Markov
chain model for the backoff window size can be represented as in
Bianchi's model \cite{Bianchi} upon exchanging $p$, the collision
probability, with the probability $P_{eq}$ defined
in~(\ref{eq.equ}).
%
%
%
%
Notice that transmission fails when
collision occurs between packets, or when data transmission is
affected by channel errors with probability $P_e$. As far as
collision is concerned, capture effects reduce the probability of
collision by an amount equal to $P_{cap}$, i.e., the probability
of capture. So, in this respect, rationales are the same as for
the 4-way handshaking mechanism discussed above.
%

Upon defining $P_{eq}$ as in~(\ref{eq.equ}), the stationary
distribution of the Bianchi's Markov chain can be evaluated by
following the same mathematical derivations dealt with in
Bianchi's model~\cite{Bianchi} whereby $P_{eq}$ is used in place
of $p$. Finally, the throughput is evaluated
through~(\ref{eq.system2}) upon solving the non-linear system of
equations in~(\ref{eq.system}).
\subsection{Markovian Process Analysis and Throughput Computation}
The objective is finding a solution of the stationary
distribution:
\[
b_{i,k}=\lim_{t\rightarrow \infty}P[s(t)=i,b(t)=k],~\forall
i\in[0,m],~\forall k\in[0,W_i-1],
\]
that is the probability of a station occupying a given state at
any discrete slot time.

First of all, note that for any $i\in [1,m)$ the following
relation holds:
\begin{equation}\label{trans_states_probabilities_2way}
\begin{array}{ll}
b_{i,0}=b_{i-1,0}P_{eq},&~\forall i\in
[1,m)\\
b_{m,0}= b_{m-1,0}P_{eq}+b_{m,0}P_{eq},&~i=m
\end{array}
\end{equation}
Proceeding backward, the following results follow by inspection:
\begin{equation}\label{collision_probabilities_2way}
\begin{array}{ll}
b_{i,0}=P_{eq}^i\cdot b_{0,0},&~\forall i\in [1,m)\\
b_{m,0}=\frac{P_{eq}^m}{1-P_{eq}}\cdot b_{0,0},&~i=m
\end{array}
\end{equation}
The stationary probabilities for any $k\in[1,W_i-1]$ follow by
resorting to the state transition diagram:
\begin{equation}\label{eq.bik_2way}
b_{i,k} = \frac{W_i-k}{W_i}\left\{
\begin{array}{ll}
(1-P_{eq})\sum_{j=0}^m b_{j,0}, &~i=0\\
P_{eq}b_{i-1,0}, &~i\in [1,m)\\
P_{eq}(b_{m-1,0}+b_{m,0}), &~ i=m
\end{array} \right.
\end{equation}
Upon utilizing the normalization condition:
%
\begin{equation}
\begin{array}{ll}
1=\sum_{i=0}^m \sum_{k=0}^{W_i-1} b_{i,k}= \frac{b_{0,0}}{2}\left[
W\left( \sum_{i=0}^{m-1}(2P_{eq})^i+
\frac{(2P_{eq})^m}{1-P_{eq}}\right) +\frac{1}{1-P_{eq}}\right]&
\end{array}
\end{equation}
the following equation for computation of $b_{0,0}$ results:
\begin{equation}
b_{0,0}=
\frac{2(1-P_{eq})(1-2P_{eq})}{(1-2P_{eq})(W+1)+P_{eq}W\left(1-(2P_{eq})^m\right)}
\end{equation}
This result is then used to compute $\tau$, the probability that a
station starts a transmission in a randomly chosen time slot. In
fact, taking into account that a data packet transmission occurs
when the backoff counter reaches zero, we have:
\begin{equation}
\begin{array}{ll}
\tau=\sum_{i=0}^m
b_{i,0}=\frac{b_{0,0}}{1-P_{eq}}=&\\
=\frac{2(1-2P_{eq})}{(1-2P_{eq})(W+1)+P_{eq}W(1-(2P_{eq})^m)}&
\end{array}
\end{equation}
The nonlinear system of equations needed to compute $\tau$,
$P_{col}$, $P_{cap}$, and $P_{eq}$ is defined as follows:
\begin{equation}\label{eq.system2way}
\left\{ \begin{array}{ll} \tau &
=\frac{2(1-2P_{eq})}{(1-2P_{eq})(W+1)+P_{eq}W(1-(2P_{eq})^m)}\\
P_{col} &= 1-(1-\tau)^{N-1} -P_{cap}\\
P_{eq}   &=
P_{col}+P_e-P_e\cdot  P_{col}\\
P_{cap}&=\sum_{i=1}^{N-1}{N \choose
i+1}\tau^{i+1}(1-\tau)^{N-i-1}\frac{1}{{(1+z_{0}g(S_{f}))}^{i}}
\end{array} \right.
\end{equation}
whereby $P_{col}$, $P_{cap}$, and $P_{eq}$ are defined as for the
4-way handshaking mechanism.

The final step in the analysis is the computation of the
normalized system throughput defined in~(\ref{eq.system2}), where
all the terms are defined as for the 4-way handshaking, excepts
for the definitions of the following time intervals for the
2-way mechanism \cite{kong}:
\begin{equation}\label{tempi_coll_error_2}
 \begin{array}{ll}
T_c       &= H+PL+ACK_{timeout}              \\
T_e       &= H+PL+ACK_{timeout} \\
T_s       &= H+PL+SIFS+ \tau_p+ACK+DIFS +\tau_p
\end{array}
\end{equation}
\section{Simulation Results and Model Validations}
\label{Simulation_results_section}
%
%
%
This section presents some simulation results for validating the
theoretical models and derivations presented in the previous
sections. We have developed a C++ simulator modelling both the DCF
protocol details in 802.11b and the backoff procedures of a
specific number of independent transmitting stations. The
simulator also takes into account all real operations of each
transmitting station, including physical propagation delays, etc.
\subsection{Simulation Setup}
Let us discuss the main functionalities of the developed
simulator. It considers an Infrastructure BSS (Basic Service Set)
with an Access Point (AP) and a pre-specified number of mobile
stations which communicates only with the AP under the hypothesis
that each station has always a packet to be transmitted, i.e.,
saturated conditions. The MAC layer is managed
by a state machine which follows the main directives specified in
the standard~\cite{standard_DCF_MAC}, namely waiting times (DIFS,
SIFS, EIFS), post-backoff, backoff, basic and RTS/CTS access mode.
As far as simulation results are concerned, we have employed MAC
layer parameters for IEEE802.11b as noted in
Table~\ref{tab.design.times}~\cite{standard_DCF_MAC}.

For the sake of simulating capture effects, the $N$ contending
stations are randomly placed in a circular area of radius $R$ (in
the simulation results presented below we assume $R=50$m), while
the AP is placed at the center of the transmission area. When two
or more station transmissions collide, the value of $\gamma$,
defined as
\begin{equation}\label{eq.defgamma}
\gamma=P_u/\sum_{k=1}^{i}P_k
\end{equation}
is evaluated for any transmitting station given their relative
distance $r_i$ from the AP. Notice that $\gamma$ in
(\ref{eq.defgamma}) is the ratio of the power $P_u$ of the useful
signal and the sum of the powers of the $i$ interfering channel
contenders transmitting simultaneously $i$ frames. Let $\gamma_j$
be the value of $\gamma$ for the $j$-th transmitting station among
the $i+1$ colliding stations. The power between a transmitter and
a receiver is the local mean power denoted $p_o$, and defined as:
\begin{equation}\label{eq.transmission_equat}
p_o=A\cdot r_i^{-n_p} P_{tx}
\end{equation}
In (\ref{eq.transmission_equat}), $n_p$ is the path-loss
exponent\footnote{In the simulated scenarios, we used the value
$n_p=3.5$.} (which is typically greater than or equal to $3.5$ in
indoor propagation conditions in the absence of the direct signal
path \cite{Rappaport}), $P_{tx}$ is the transmitted power, and
$A\cdot r_i^{-n_p}$ is the deterministic path-loss
\cite{Rappaport}. Both $A$ and $P_{tx}$ are identical for all
transmitted frames. When signal transmission is affected by
Rayleigh fading, the instantaneous power of the signal received by
the receiver placed at a mutual distance $r_i$ from the
transmitter is exponentially distributed as:
\[
f(x)=\frac{1}{p_o}e^{-\frac{x}{p_o}}, x>0
\]
The values of $\gamma_j$ for each colliding station are compared
with the threshold $z_o\cdot g(S_f)$. Then, the transmitting
station for which $\gamma_j$ is above the threshold captures the
channel.

Throughput simulations are accomplished by averaging over $100$
sample scenarios, whereby any transmitting scenario considers a
set of $N$ randomly distributed (with a uniform probability
mass function) stations over a circular area of radius 50 m as
specified above.

The physical layer (PHY) of the basic 802.11b standard is based on
the spread spectrum technology. Two options are specified, the
Frequency Hopped Spread Spectrum (FHSS) and the DSSS (Direct
Sequence Spread Spectrum). The FHSS uses Frequency Shift Keying
(FSK) while the DSSS uses Differential Phase Shift Keying (DPSK)
or Complementary Code Keying (CCK). The 802.11b employs DSSS at
various rates including one employing CCK encoding 4 and 8 bits on
one CCK symbol. The four supported data rates in 802.11b are 1, 2,
5.5 and 11 Mbps.
%

The FER as a function of the SNR can be computed as follows:
\begin{equation}\label{fer_1}\small
P_e(SNR)=1-\left[1-P_e(PLCP,SNR)\right]\cdot
\left[1-P_e(DATA,SNR)\right]
\end{equation}
where,
\begin{equation}\label{fer_2}\small
P_e(PLCP,SNR)=1-\left[1-P_b(BPSK,SNR)\right]^{8\times PLCP},
\end{equation}
and
\begin{equation}\label{fer_3}\small
P_e(DATA,SNR)=1-\left[1-P_b(TYPE,SNR)\right]^{8\times (DATA+MAC)}.
\end{equation}
$P_b(BPSK,SNR)$ is the BER as a function of SNR for the lowest
data transmit rate employing DBPSK modulation, DATA denotes the
data block size in bytes, and any other constant byte size in
above expression represents overhead. Note that the FER,
$P_e(SNR)$, implicitly depends on the modulation format used.
Hence, for each supported rate, one curve for $P_e(SNR)$ as a
function of SNR can be generated. $P_b(TYPE,SNR)$ is modulation
dependent whereby the parameter $TYPE$ can be any of the following
$TYPE\in \{DBPSK,DQPSK,CCK5.5,CCK11\}$\footnote{The acronyms are
short for Differential Binary Phase Shift Keying, Differential
Quadrature Phase Shift Keying and Complementary Code Keying,
respectively.}.

For DBPSK and DQPSK modulation formats, $P_b(TYPE,SNR)$ can be
well approximated by \cite{SimonAlouini}:
\begin{equation}\small
\label{ber_function}
\frac{2}{\max(\log_{2}M,2)}\sum_{i=1}^{\max(\frac{M}{4},1)}\frac{1}{\pi}\int_{0}^{\frac{\pi}{2}}
\frac{1}{1+\gamma\frac{1}{\sin^2\theta}\log_{2}M\sin^2\left(\frac{(2i-1)\pi}{M}\right)}d\theta
\end{equation}
whereby, $M$ is the number of bits for modulated symbols, $\gamma$
is the signal-to-noise ratio, and $\theta$ is the signal direction
over the Rayleigh fading channel.

In so far as the computation of the FER is concerned, it should be
noted that data packet error rates of different contending
stations differ. For simplicity, we assume that data packets
transmitted by different stations are affected by the same FER. Of
course, this is a simplifying assumption which is indeed justified
by the following pragmatic considerations. We have verified by
simulation that the aggregate throughput is quite insensitive on
the fact that different stations experience different FER values
on the transmitted packets, provided that the maximum FER value
affecting the packets of whatever station in the network, is lower
than\footnote{Notice that $8\times 10^{-2}$ corresponds, for
$E[PL]=1024$ bytes, to the worst allowed FER associated to the
receiver sensitivity specified in the IEEE 802.11b standard
\cite{standard_DCF_MAC}.} $8\times 10^{-2}$. As a reference
scenario, we considered $N=9$ contending stations divided
uniformly in three groups, with the assumption that the stations
in each group transmit packets affected by the same FERs noted in
the first column of Table \ref{tab.throughput_per}. Then, we
evaluated through simulation the aggregate throughput, shown in
the second column, in six configurations, three of which are
characterized by a maximum FER equal to $10^{-2}$, while the
others are associated to a maximum FER equal to $10^{-3}$. The
other network parameters are noted in the next section. Upon
comparing the scenario in which all the stations transmit packets
affected by the same FER, with a scenario whereby some packets are
affected by lower FER values, it is possible to note that the
obtained throughput does not change significantly, confirming the
simplifying assumption considered in this paper. Furthermore, this
is a common hypothesis widely used in the literature
\cite{Chatzimisios}.

Channel errors on the transmitted packets have been accounted for
as it is done within ns-2~\cite{xiuchao} simulator. In other words, a
uniformly distributed binary random variable $X_e$ is generated in
order to decide if a transmitted packet is received erroneously.
The statistic of such a random variable is $P(X_e=1)=P_e(SNR)$ (as
specified in (\ref{fer_1})), and $P(X_e=0)=1-P_e(SNR)$.
%
%
%
%
\subsection{Simulation Results}
In what follows, we shall present theoretical and simulation
results for the lowest supported data rate. We note that by
repeating the process, similar curves can be generated for all
other types of modulation formats. All we need is really the BER
as a function of SNR for each modulation format and the
corresponding raw data rate over the channel. If the terminals use
rate adaptation, then under optimal operating condition, the
achievable throughput for a given SNR is the maximum over the set
of modulation formats supported.

In the following, we present simulation results for the raw data
rate of 1 Mbps. We have verified a close match between theoretical
and simulated performance for other transmitting data rates as
well. In the simulation results presented below we assume the
following values for the contention window: CWmin=32, m=5, and
CWmax = $2^m\cdot$CWmin = 1024.

The theoretical behavior of the probabilities $\tau$, $P_{eq}$,
and $P_{col}$ are depicted in Figs.~\ref{N20_tau_1024bytes}
and~\ref{N20_P_eq_1024bytes} for both 2 way and 4 way hand-shaking
mechanisms as a function of the channel SNR. All the curves have
been drawn for $N=20$ contending stations and for a payload size
equal to 1024 bytes. The curves have been parameterized with respect to
the capture threshold $z_0$. A comparative analysis of the two
figures reveals that for increasing values of the quality of the
channel, as exemplified by the SNR, the transmission probability
$\tau$ increases up to reaching a saturated level around $40$dB,
above which the channel quality can be well assumed as ideal.
Furthermore, this probability increases for lower values of the
capture threshold $z_0$, i.e., for higher values of the capture
probability.

A comparative analysis of the curves shown in
Fig.~\ref{N20_P_eq_1024bytes} reveals that the equivalent
transmission failure probability decreases for increasing values
of SNR, achieving asymptotic values essentially equal to collision
probability at high SNR values. Notice that the relation defining
probability $\tau$ for the 2 way mechanism, as defined in
(\ref{eq.tau}), tends to the relation found by Bianchi
\cite{Bianchi}:
\[
\tau=\frac{2(1-2p)}{(1-2p)(W+1)+pW(1-(2p)^m)}
\]
when $SNR\rightarrow\infty$, i.e., for $P_e(SNR)\rightarrow
0$, and $P_{cap}\rightarrow 0$. In this case,
$P_{eq}\rightarrow p$, where $p$ is the collision probability as
defined by Bianchi, and here identified by $P_{col}$. In this
respect, our model is more general and also embraces Bianchi's
model as an asymptotic case. This is clearly highlighted in
Figs.~\ref{N20_tau_1024bytes} and~\ref{N20_P_eq_1024bytes} whereby
the Bianchi's probabilities as specified in \cite{Bianchi} are
depicted as horizontal lines due to the independence of the
Bianchi's model on both capture effects and channel errors. On
the other hand, for very low values of SNR, the equivalent failure
probability $P_{eq}$ is practically 1 no matter what values are
assumed by the other parameters.
%

Fig.~\ref{2W_1024b_N5} shows the behavior of the saturation
throughput for the 2 way mechanism as a function of the SNR, for
three different capture thresholds, $z_0$, and for 5 transmitting
stations. Simulated points are marked by star on the respective
theoretical curves. As expected, throughput improves as the SNR
increases up to 35 dB. Furthermore, throughput improves because of
the capture effects.

Fig.~\ref{2W_1024b_N20} shows the behavior of the saturation
throughput for the 2 way mechanism as a function of the SNR, for
three different capture thresholds, $z_0$, and for 20 transmitting
stations. Upon comparing the curves shown in
Fig.s~\ref{2W_1024b_N5} and~\ref{2W_1024b_N20}, it is easily seen
that capture effects allow the system throughput to be almost the
same independently from the number of stations, since throughput
curves related to $z_0=1$ dB are very close. On the other hand,
when capture probability is low, i.e., for $z_0=24$dB, collision
probability reduces the throughput as the number of
contending stations increase.

With the aim of comparing the throughput predicted by the proposed
model in a variety of transmission conditions to the Bianchi's
throughput, Fig.~\ref{2W_1024b_N20} also shows the Bianchi's
throughput along with the maximum throughput $S_m$ achievable upon
optimizing the minimum contention window, $W$, as discussed in
\cite{Bianchi}. The maximum throughput evaluated by Bianchi is
defined as:
\begin{equation}\label{Smax_bianchi}
S_m=\frac{E[PL]}{T_s+\sigma K+T_c\left(K(e^{1/K}-1)-1\right)}
\end{equation}
whereby $K=\sqrt{\frac{T_c}{2\sigma}}$, while $T_c$ and $T_s$ are
defined in (\ref{tempi_coll_error_2}) for the 2-way mechanism, and
(\ref{tempi_coll_error}) for the 4-way mechanism. Notice that, as
already suggested by Bianchi, $S_m$ is independent of the number
$N$ of contending stations.

First of all, notice that as $SNR\rightarrow\infty$ and in absence
of capture, the throughput predicted by our model tends to the
Bianchi's throughput, which also corresponds to about $z_0=24$dB,
i.e., a transmission scenario in which capture threshold is so
high that capture probability is very low. On the other hand, for
decreasing values of $z_0$, throughput tends to increase, as
expected, in the presence of capture. Secondly, for values of
$z_0=6$dB, capture effects are such that the throughput reaches
the value $S_m$ predicted by Bianchi under the hypothesis of
employing an optimal minimum contention window size, even though
in our simulation setup the minimum contention window is not
chosen to be the optimal value yielding (\ref{Smax_bianchi}). For
increasing values of capture probabilities, as exemplified by
$z_0=1$dB, throughput goes above $S_m$. This is essentially due to
the fact that capture tends to reduce the collision probability
experienced by the stations which attempt to access the channel simultaneously,
validating the key hypothesis suggested in
(\ref{eq.col}). Indeed, simulation results confirm such a
hypothesis.
%
%

In order to assess throughput performance as a function of the
payload size, Fig.~\ref{2W_different_payl_sizes} shows throughput
performance as a function of the SNR, for two different values of
the payload sizes. Depending on the channel quality as exemplified
by the SNR on the abscissa in Fig.~\ref{2W_different_payl_sizes},
it could be preferable to transmit shorter packets for low SNRs.
On the other hand, when channel quality improves, longer packets
allow one to increase throughput.
%

Finally, Fig.s~\ref{sim2W_1024byte} and~\ref{sim4W_1024byte} show
the behavior of the saturation throughput as a function of the
number of the contending stations for 2 and 4 way mechanism,
respectively, for various key values of the parameters SNR and
$z_0$. In both figures, SNR$=\infty$ is used to signify ideal
transmission (i.e., $P_e=0$), while $z_0=\infty$ means absence of
capture effects. Notice that, as expected, in the 4 way mechanism,
saturation throughput performance is less sensitive to capture
effects. For comparison purposes, Fig.s~\ref{sim2W_1024byte} shows
Bianchi's throughput along with the maximum achievable
throughput in (\ref{Smax_bianchi}). Notice that Bianchi's
throughput is superimposed on the curve labelled $SNR=\infty$ and
$z_0=\infty$, as already noted from the simulation results shown
in the previous figures. Achievable throughput in a variety of
transmission conditions is usually less than $S_m$ in
(\ref{Smax_bianchi}), specially when $SNR$ is lower than $45$dB:
in this case, channel errors tend to dominate reducing the
throughput. On the other hand, throughput can be higher than $S_m$
for a low number of contending stations (in the considered
scenario, $N\le 20$ contending stations) and for low values
of capture thresholds $z_0$, say $z_0\le 6$dB. On the other hand,
employing the 4 way mechanism, throughput is very close to the
theoretical Bianchi's value $S_m=0.86$ Mbps no matter what the capture
threshold $z_0$. This is due to the impact of the 4 way
mechanism mitigating the effects of collisions between
multiple contending stations.
%
\section{Conclusions}
In this paper, we provided an extension of the Markov model
characterizing the DCF behavior at the MAC layer of the IEEE802.11
series of standards accounting for channel induced errors and
capture effects typical of fading environments. The modelling
allows taking into consideration the impact of channel contention
in throughput analysis which is often not considered or it is
considered in a static mode by using a mean contention period.
Subsequently, based on justifiable assumptions, the stationary
probability of the Markov chain is calculated to obtain the
saturation throughput. The closed form expressions allow
derivation of the throughput as a function of a multitude of
system level parameters including packet and header sizes for a
variety of applications. Simulation results confirm the validity
of the proposed theoretical models.
%
%
\clearpage
\clearpage
\figuramedia{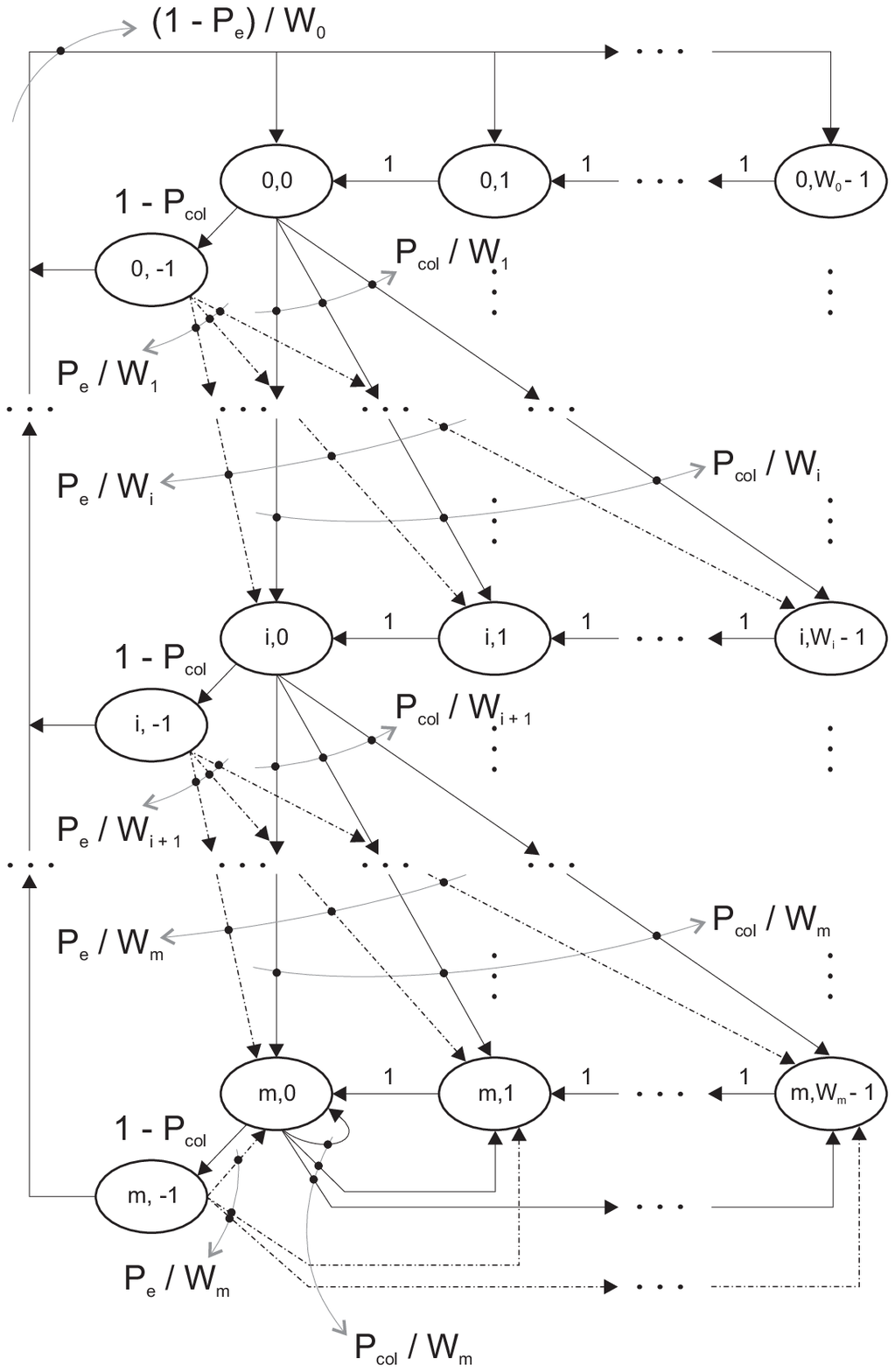}{Markov chain for the contention model
based on the 4-way handshaking technique, considering the effects
of capture and channel induced errors.}{fig.chain}
\clearpage
\figura{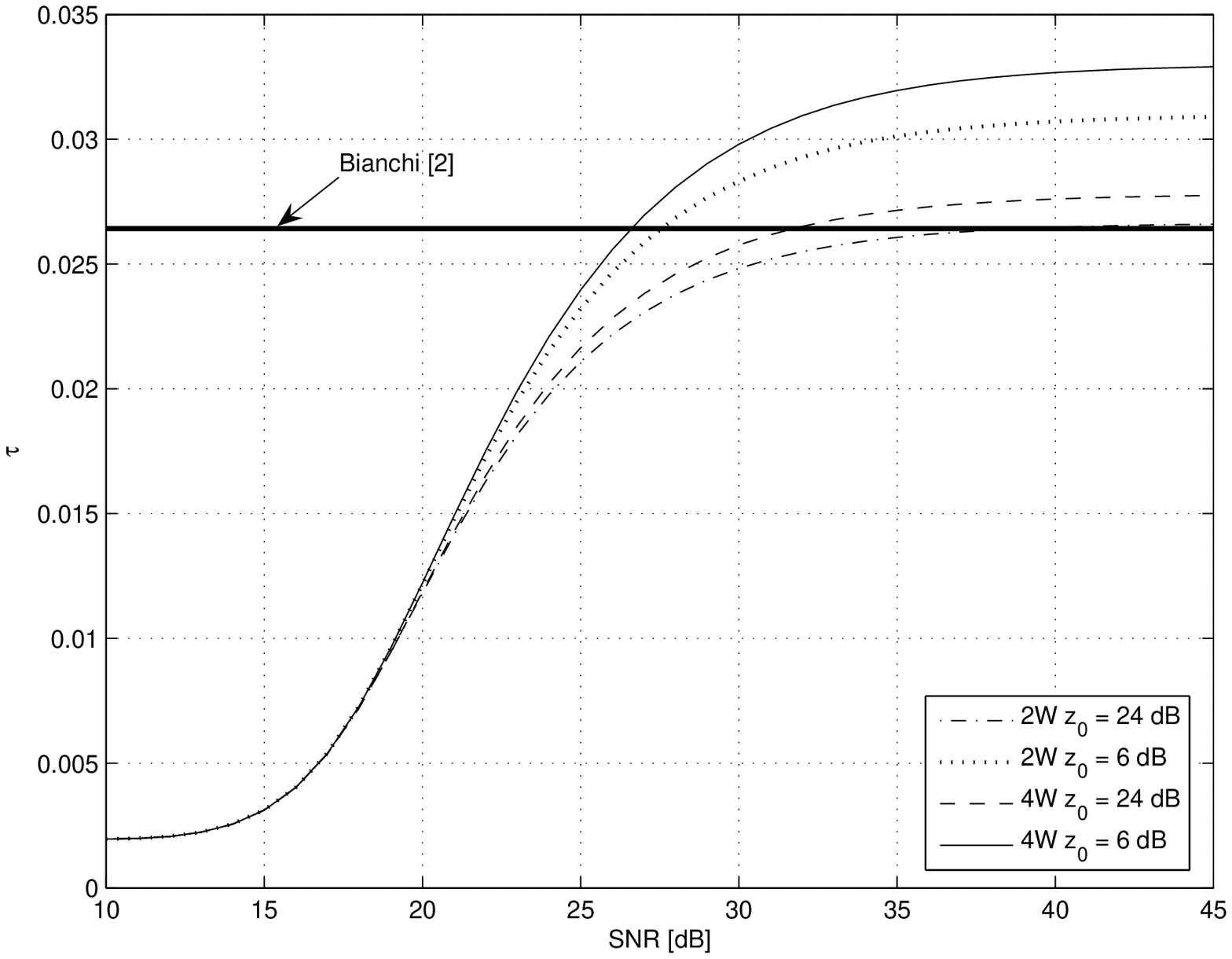}{Theoretical behavior of the
probability $\tau$ as a function of SNR for both 2 way (curves
labelled 2W) and 4 way (curves labelled 4W) handshaking
mechanisms. Curves have been obtained for the capture threshold
$z_0$, payload size equal to 1024 bytes and for $N=20$ contending
stations.}{N20_tau_1024bytes}
\clearpage
\figura{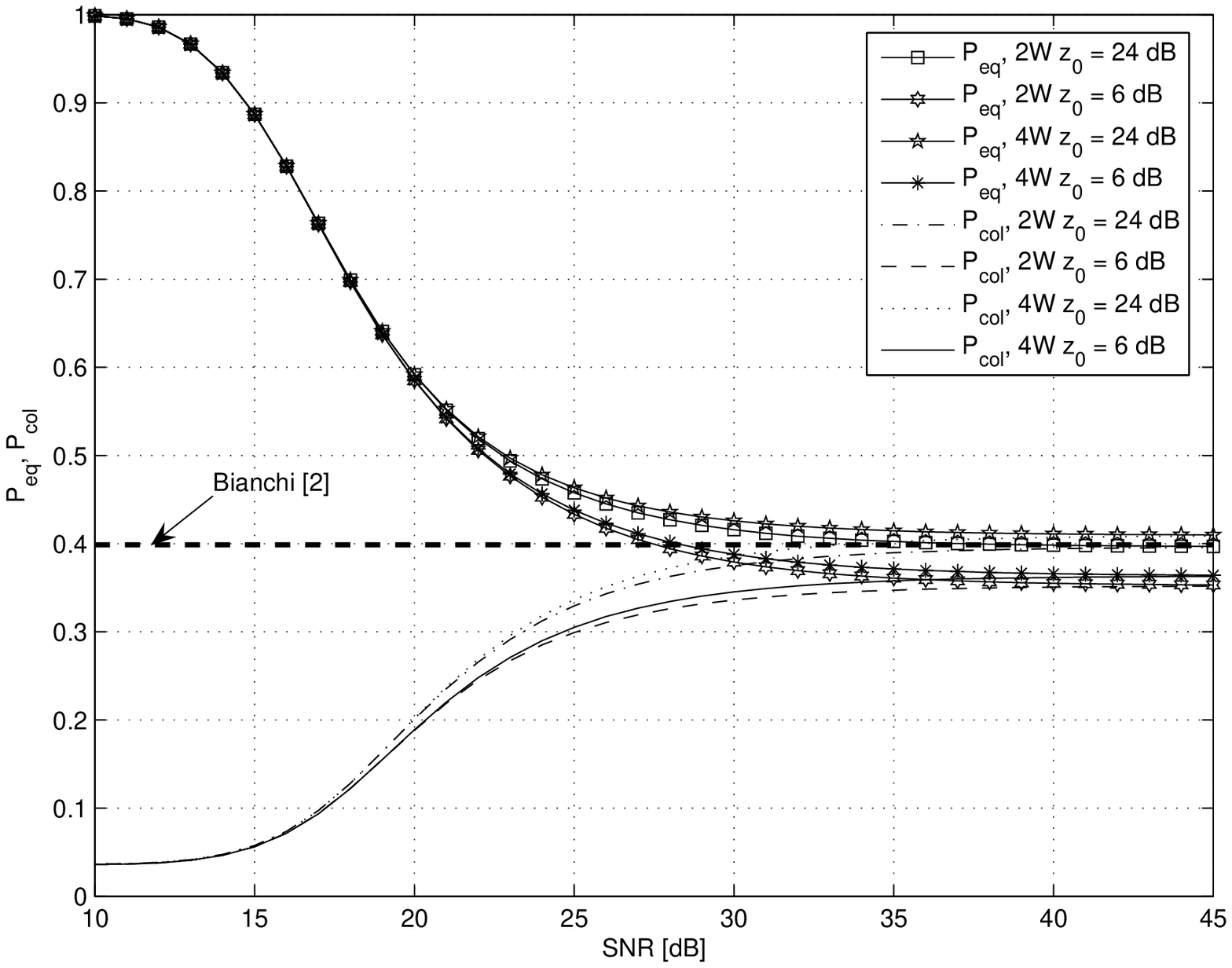}{Theoretical behaviour of the
equivalent failure transmission probability $P_{eq}$ as a function
of SNR for both 2 way (curves labelled 2W) and 4 way (curves
labelled 4W) handshaking mechanisms. Curves have been obtained for
the capture threshold $z_0$, payload size equal to 1024 bytes and
for $N=20$ contending stations.}{N20_P_eq_1024bytes}
\clearpage
\figura{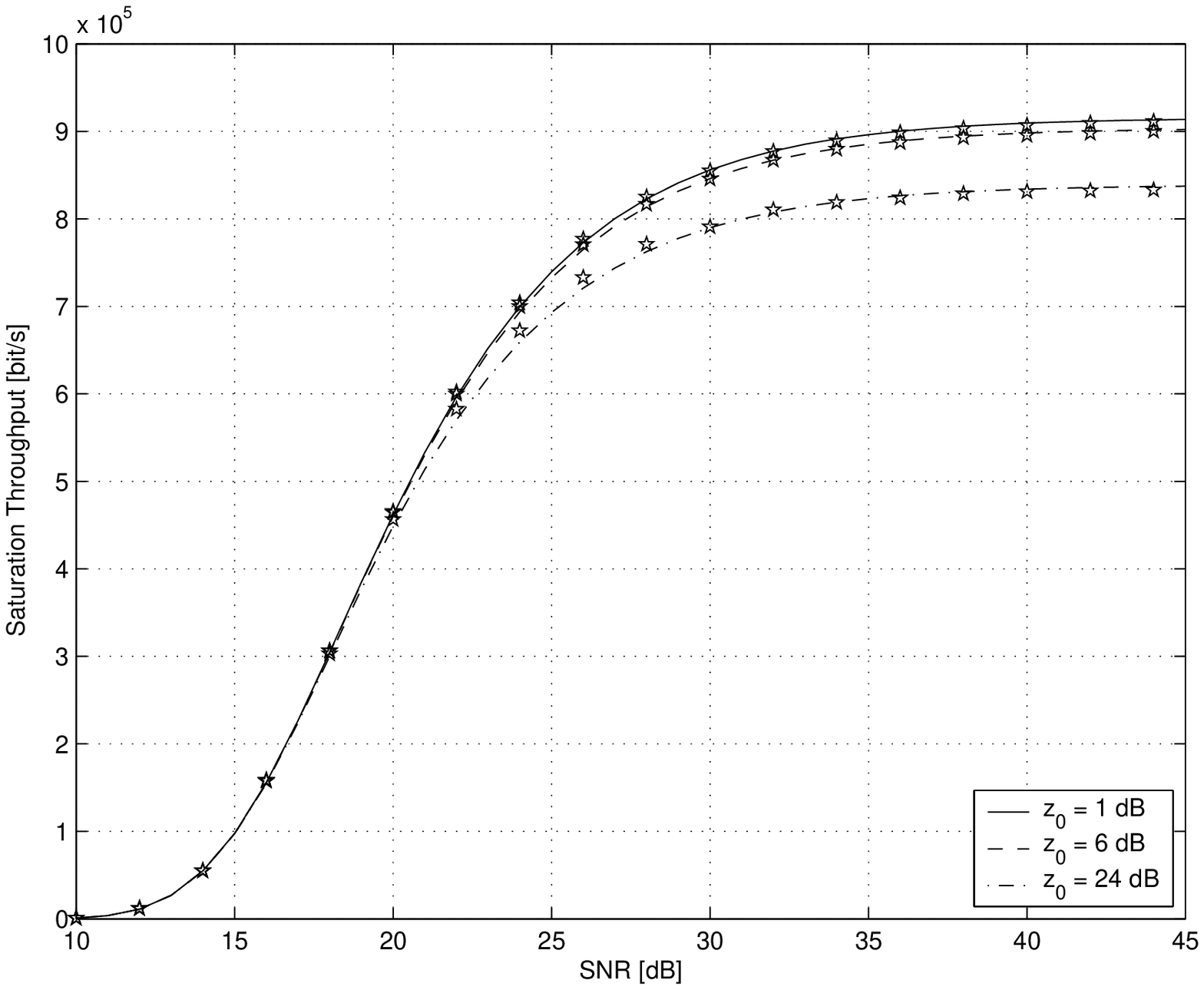}{Saturation throughput for the 2 way
mechanism as a function of the SNR, for three different capture
thresholds $z_0$. Simulated points are identified by star-markers
over the respective theoretical curves. Payload size is 1024
bytes, while the number of contending stations is
$N=5$.}{2W_1024b_N5}
\clearpage
\figura{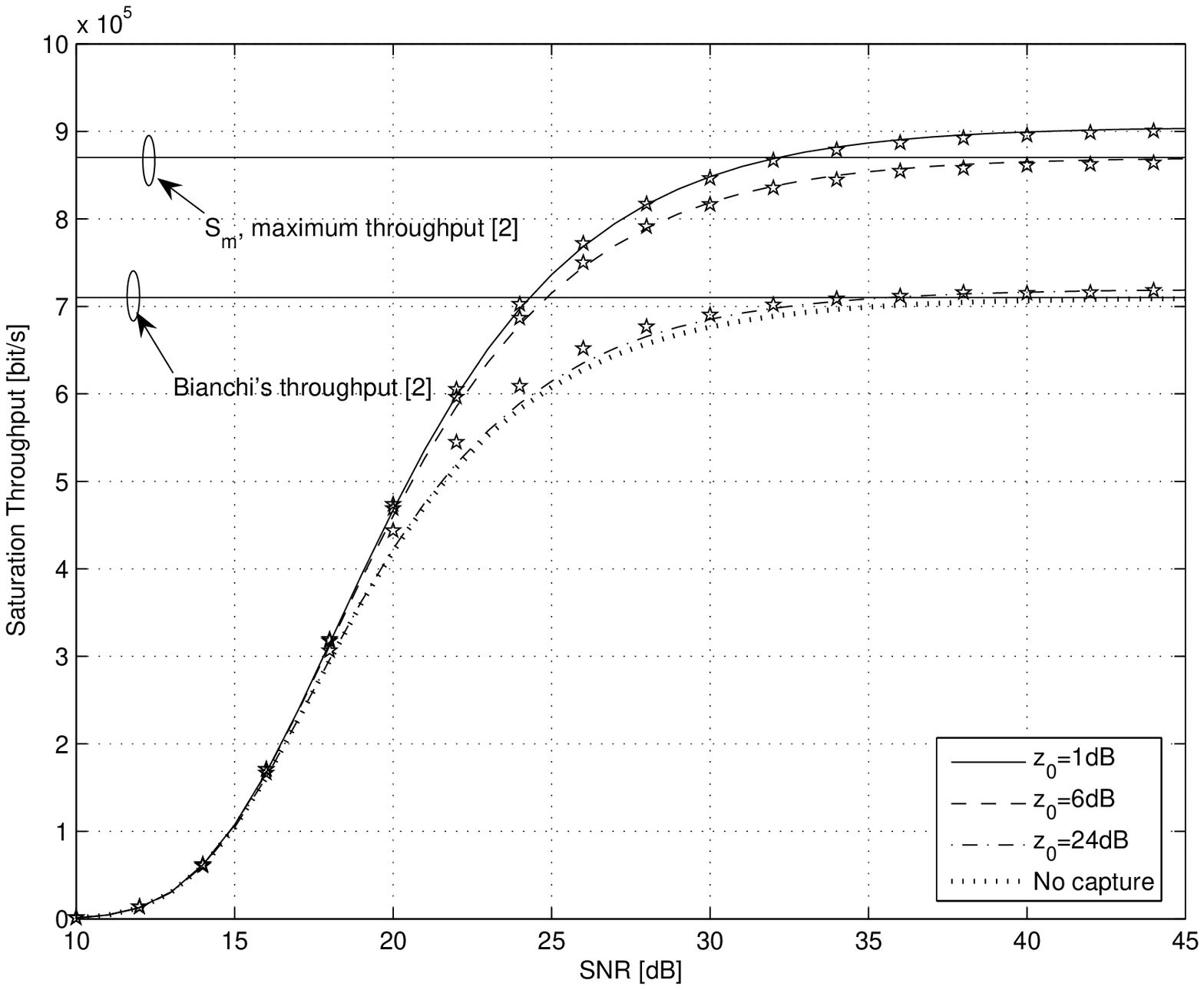}{Saturation throughput for the 2 way
mechanism as a function of the SNR, for three different capture
thresholds $z_0$. Simulated points are identified by star-markers
over the respective theoretical curves. Payload size is 1024
bytes, while the number of contending stations is
$N=20$.}{2W_1024b_N20}
\clearpage
\figura{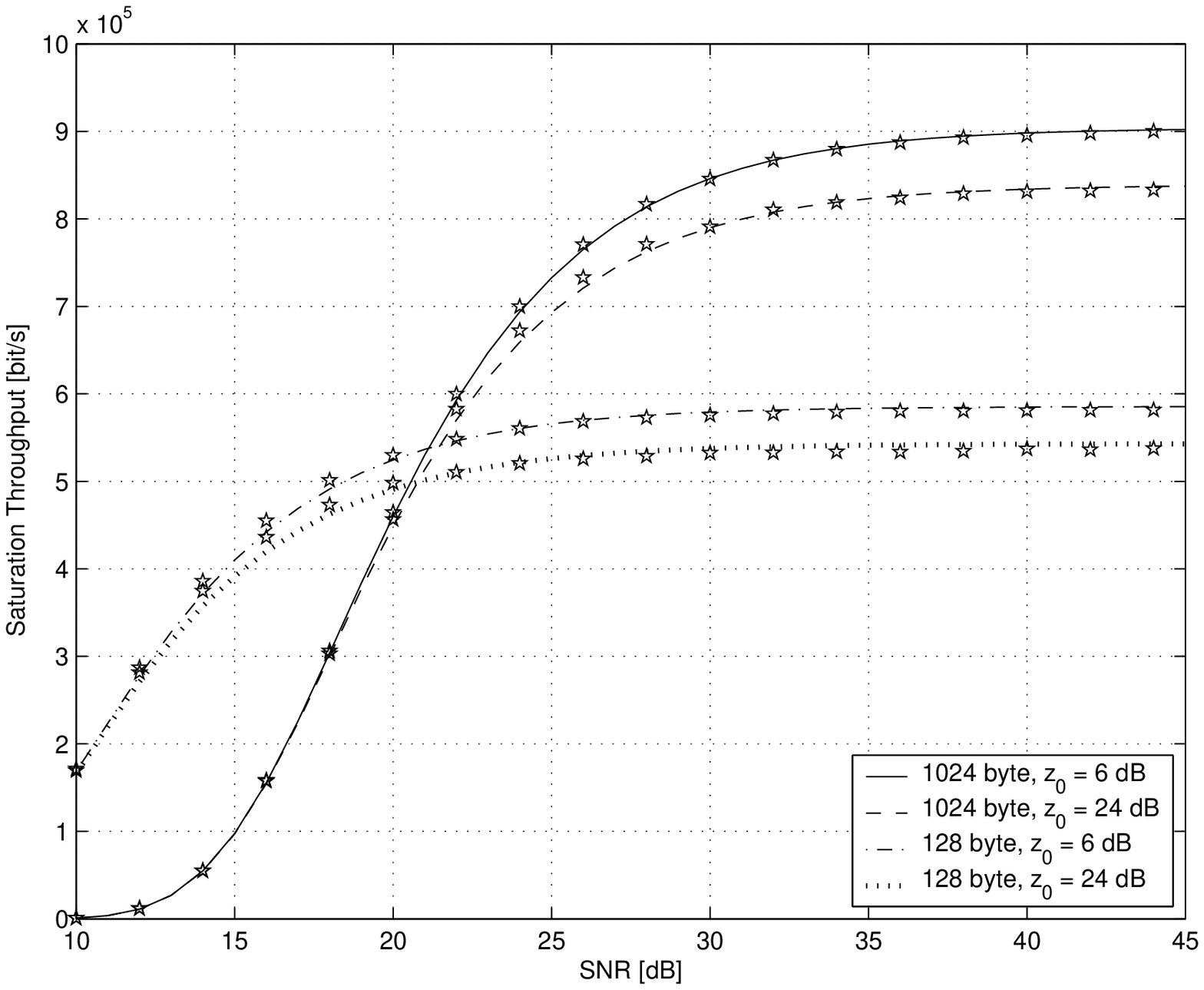}{Saturation throughput for the
2 way mechanism as a function of the SNR, for payload sizes 128
and 1024 bytes, and $z_0=6,~24$~dB. Simulated points are
identified by star-markers over the respective theoretical
curves.}{2W_different_payl_sizes}
\clearpage
\figura{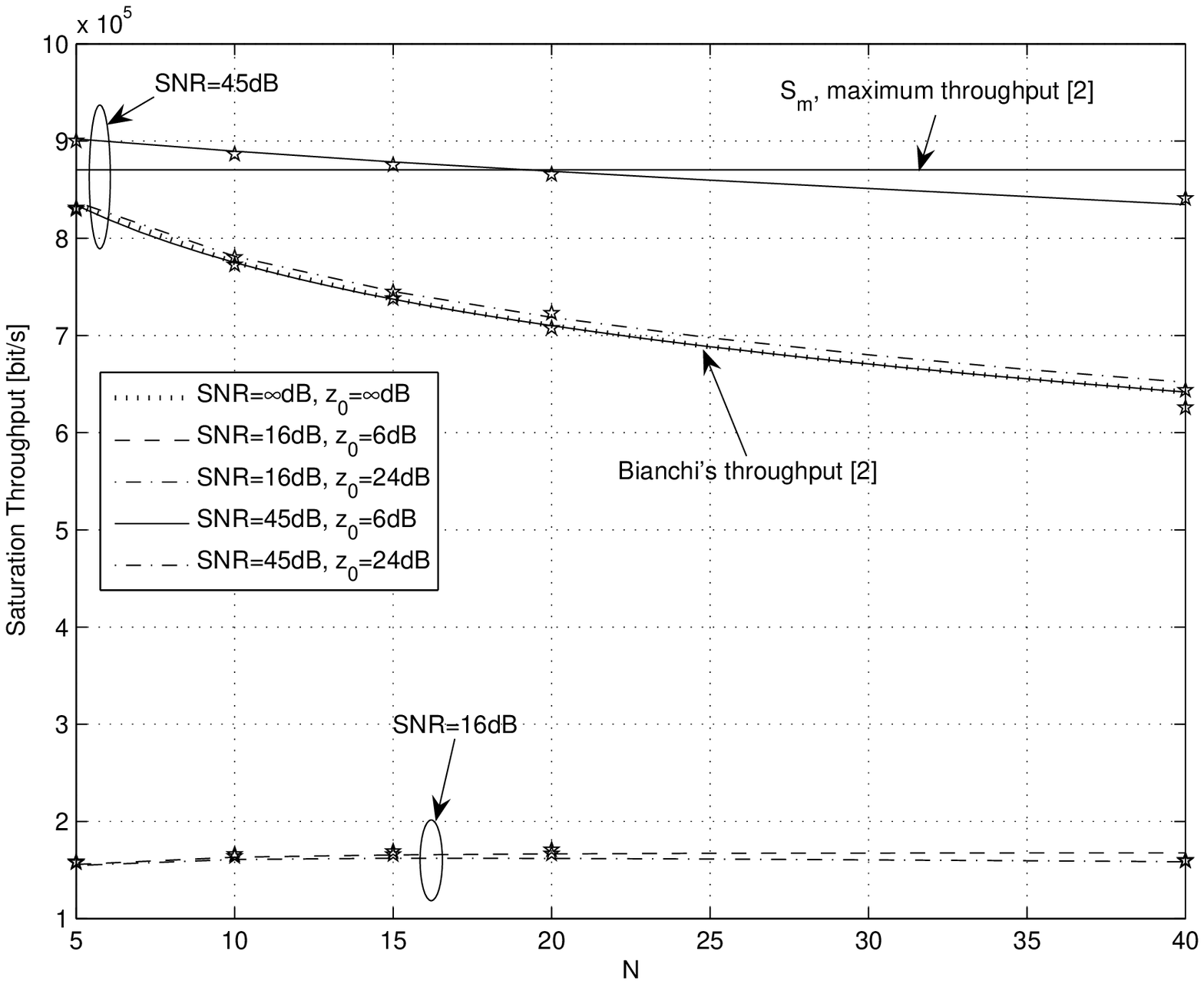}{Saturation throughput for the 2 way
mechanism as a function of the number of contending stations, for
various values of the SNR and $z_0$ parameters. Payload size is
1024 bytes. Simulated points are identified by star-markers over
the respective theoretical curves.}{sim2W_1024byte}
\clearpage
\figura{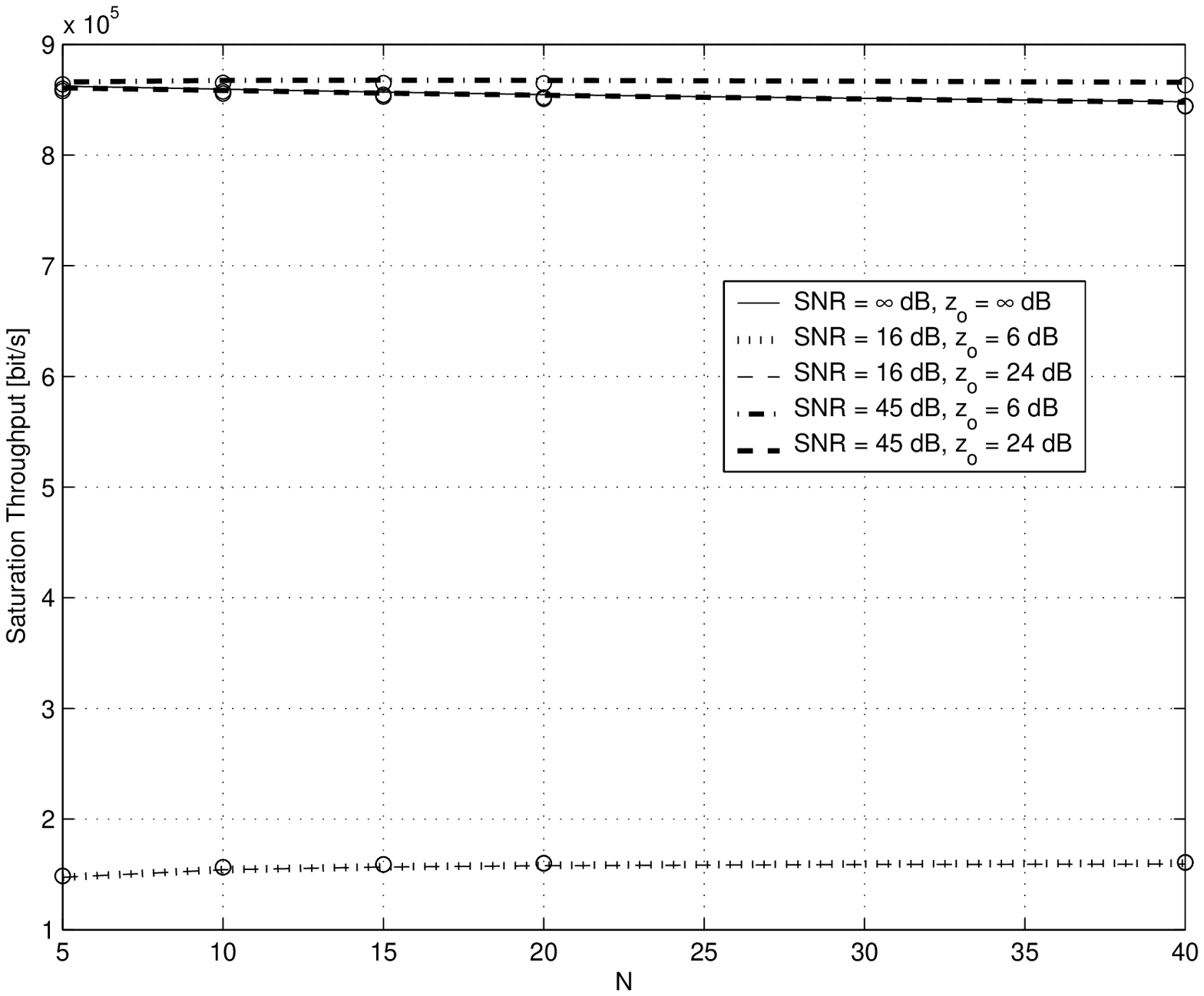}{Saturation throughput for the 4 way
mechanism as a function of the number of contending stations, for
various values of the SNR and $z_0$ parameters. Payload size is
1024 bytes. Simulated points are identified by star-markers over
the respective theoretical curves.}{sim4W_1024byte}
\clearpage
\begin{table}\caption{Typical network parameters}
\begin{center}
\begin{tabular}{|c|c|}\hline
\hline MAC header & 24 bytes\\
\hline PHY header & 16 bytes\\
\hline Payload size & 1024 bytes\\
\hline ACK & 14 bytes\\
\hline RTS & 20 bytes\\
\hline CTS & 14 bytes\\
\hline
\hline propagation delay $\tau_p$ & 1 $\mu s$\\
\hline Slot time & 20 $\mu s$\\
\hline SIFS & 10 $\mu s$\\
\hline DIFS & 50 $\mu s$\\
\hline EIFS & 300 $\mu s$\\

\hline ACK timeout & 300 $\mu s$\\
\hline CTS timeout & 300 $\mu s$\\
\hline\hline
\end{tabular}
 \label{tab.design.times}
\end{center}
\end{table}
\clearpage
\begin{table}\caption{Throughput dependence on the per-station FER in a scenario with $N=9$
contending stations using the 2-way mechanism without capture.
Packet size is $E[PL]=1024$ bytes.}
\begin{center}
\begin{tabular}{c|c}
\hline \hline  FER & $S_i$~[Mbps]\\ \hline \hline
 $10^{-2}, 10^{-2}, 10^{-2}$ & 0.777\\
\hline
 $10^{-2}, 10^{-3}, 10^{-4}$ & 0.781 \\
\hline
$10^{-2}, 10^{-4}, 10^{-5}$ & 0.781 \\
\hline \hline
 $10^{-3}, 10^{-3}, 10^{-3}$ & 0.784  \\
\hline
 $10^{-3}, 10^{-4}, 10^{-5}$ & 0.786 \\
\hline
 $10^{-3}, 10^{-5}, 10^{-6}$ & 0.785 \\
\hline \hline
\end{tabular}
 \label{tab.throughput_per}
\end{center}
\end{table}
\end{document}